\documentclass[12pt]{article}

\usepackage{prooftree}

\usepackage{amssymb}

\def\ruleone#1#2{\prooftree #1 \justifies #2 \endprooftree}
\def\ruletwo#1#2#3{\ruleone{#1\quad #2}{#3}}
\def\ruledot#1{\prooftree \proofdotseparation=1.2ex\proofdotnumber=3
               \leadsto#1 \justifies #1 \endprooftree}

\def\imp#1#2{#1\rightarrow_p #2}
\def\vp#1#2{#1 \vee_p #2}

\def\enc{{\sl enc}}
\def\dec{{\sl dec}}
\def\ff{{\mathbf{f}}}
\def\crack{{\sl crack}}
\def\gr{{\sl gr}}

\newtheorem{theo}{Theorem}[section]
\newtheorem{lem}[theo]{Lemma}

\newtheorem{defin}[theo]{Definition}

\def\proof{\par\smallskip\noindent {\bf Proof. }}
\def\com{\par\medskip\noindent{\bf Comment. }}
\def\fin{\hfill\rule{1ex}{1ex}\\[2pt]}

\title{Primal implication as encryption}
\author{Vladimir N. Krupski}
\begin{document}
\maketitle

\begin{abstract}
We propose a ``cryptographic'' interpretation for the
propositional connectives of primal infon logic introduced by
Y.~Gurevich and I.~Neeman and prove the corresponding soundness
and completeness results. Primal implication $\imp{\varphi}{\psi}$
corresponds to the encryption of $\psi$ with a secret key
$\varphi$, primal disjunction $\vp{\varphi}{\psi}$ is a group key
and $\bot$ reflects some backdoor constructions such as full
superuser permissions or a universal decryption key. For the logic
of $\bot$ as a universal key (it was never considered before) we
prove that the derivability problem has linear time complexity. We
also show that the universal key can be emulated using primal
disjunction.
\end{abstract}

\section{Introduction}
Primal Infon Logic (\cite{GurNee2008}, \cite{GurNee2009},
\cite{GurNee2011}, \cite{BekGur2011}, \cite{CotGur2012})
formalizes the concept of {\em infon}, i.e. a message as a piece
of information. The corresponding derivability statement
$\Gamma\vdash \varphi$ means that the principal can get (by herself,
without any communication) the information $\varphi$ provided she
already has all infons $\psi\in\Gamma$.

{\em Primal implication } $(\imp{}{})$ that is used in Primal
Infon Logic to represent the conditional information is a
restricted form of intuitionistic implication defined by the
following inference rules:

$$
\ruleone{\Gamma\vdash \psi}{\Gamma\vdash \imp{\varphi}{\psi} \using
(\imp{}{}I)} \quad,\quad \ruletwo{\Gamma\vdash \varphi}{\Gamma\vdash
\imp{\varphi}{\psi}}{\Gamma\vdash \psi\using (\imp{}{}E)} \;.
$$

These rules admit cryptographic interpretation of primal
implication $\imp{\varphi}{\psi}$ as some kind of digital envelop:
it is an infon, containing the information $\psi$ encrypted by a
symmetric key (generated from) $\varphi$. Indeed, the introduction
rule $(\imp{}{}I)$ allows to encrypt any available message  by any
key. Similarly, the elimination rule $(\imp{}{}E)$ allows to
extract the information from the ciphertext provided the key is
also available. So the infon logic incorporated into communication
protocols (\cite{GurNee2008}, \cite{GurNee2009}) is a natural tool
for manipulating with commitment schemes (see \cite{Goldr})
without detailed analysis of the scheme itself.

\par\medskip\noindent
{\bf Example.} (cf. \cite{Blum}). Alice and Bob live in different
places and  communicate via a telephone line or by e-mail. They
wish to play the following game distantly. Each of them picks a
bit, randomly or somehow else. If the bits coincide then Alice
wins; otherwise Bob wins. Both of them decide to play fair but
don't believe in the fairness of the opponent. To play fair means
that they honestly declare their choice of a bit, independently of
what the other player said. So they use cryptography.

We discuss the symmetric version of the coin flipping protocol
from \cite{Blum} in order to make the policies of both players the
same. Consider the policy of one player, say Alice. Her initial
state can be represented by the context
$$
\Gamma=\{A\,said\;m_a,\;A\,said\;k_a,\;A\,IsTrustedOn\;m_a,\;A\,IsTrustedOn\;k_a\},
$$
where infons $m_a$ and $k_a$ represent the chosen bit and the key
Alice intends to use for encryption.
%One can use a random
%bit as a key here. In such a case, the infon $\imp{k_a}{m_a}$ will
%denote the ciphertext $(m_b+k_b)\,mod\;2$.
%The set of infons involved in this example is finite. One can
%assume that all of them are translated into binary strings of the
%same length by some hash function and use bitwise {\sl XOR}
%function as the encryption. See Appendix for more details.  }.
 Her choice is recorded by
infons $A\,said\;m_a$ and $A\,said\;k_a$ where $A\,said$ is the
quotation modality governed by the modal logic {\bf K}.\footnote{
The only modal inference rule that is used in this paper is
$X\,said\; \varphi, \,X\,said\,(\imp{\varphi}{\psi}) \vdash
X\,said\; \psi$. It is admissible in {\bf K}. For more details
about modalities in the infon logic see
\cite{BekGur2011},\cite{CotGur2012}.} Alice simply says, to
herself, the infons $m_a$ and $k_a$.

The remaining two members of $\Gamma$ reflect the decision to play
fair. The infon $X\,IsTrustedOn\;y$ abbreviates $\imp
{(X\;said\;y)}{y}$. It provides the ability to obtain the actual
value of $y$ from the declaration $X\;said\;y$, so Alice can
deduce the actual $m_a$ and $k_a$ she has spoken about.

The commit phase. Alice derives $m_a$ and $\imp {k_a}{m_a}$ from
her context by rules $(\imp{}{}E)$, $(\imp{}{}I)$ and sends the
infon $\imp {k_a}{m_a}$ to Bob. Bob acts similarly, so Alice will
receive a message from him and her context will be extended to
$$
\Gamma'=\Gamma\cup\{B\,said\;(\imp {k_b}{m_b})\}.
$$

The reveal phase. After updating the context Alice obtains $k_a$
by rule $(\imp{}{}E)$ and sends it to Bob. He does the same, so
Alice's context will be
$$
\Gamma''=\Gamma'\cup\{\;B\;said\;k_b\}.
$$
Now by reasoning in {\bf K} Alice deduces $B\;said\;m_b$. She also
has $A\;said\;m_a$, so it is clear to her  who wins. Alice simply
compares these infons with the patterns $B\;said\;0$, $B\;said\;1$
and $A\;said\;0$, $A\;said\;1$ respectively.

The standard analysis of the protocol shows that  Bob will come to
the same conclusion. Moreover, Alice can be sure that she is not
cheated provided she successively  follows her policy up to the
end.\footnote{ Here we suppose that the encryption method is
practically strong and unambiguous.
%Any ciphertext corresponds to the unique plaintext.
It is impossible for a player who does not know the encryption key
to restore the plaintext from a ciphertext. It is also impossible
for him to generate two key-message pairs with different messages
and the same ciphertext. } The same with Bob.

Note that infon logic is used here as a part of the protocol. It
is one of the tools that provide the correctness. But it does not
prove the correctness. In order to formalize and prove the
correctness of protocols one should use much more powerful formal
systems. \fin

We make our observation precise by defining interpretations of
purely propositional part of infon logic in ``cryptographic''
infon algebras and proving the corresponding soundness and
completeness theorems.

In Section \ref{SEMANTICS} this is done for the system {\bf P}
which is the $\{\top,\wedge,\imp{}{}\}$-fragment of infon logic.
We also show that the quasi-boolean semantics for {\bf P} (see
\cite{BekGur2011}) is essentially a special case of our semantics.

In Section \ref{CONSTANTBOT} we show that  $\bot$ can be used to
reflect some backdoor constructions.  Two variants are considered:
system ${\bf P}[\bot]$ from \cite{BekGur2011} with the usual
elimination rule for $\bot$ and a new system ${\bf P}[\bot_w]$
with a weak form of elimination rule for $\bot$. The first one
treats $\bot$ as a root password, and the second one --- as a
universal key for decryption. For almost all propositional primal
infon logics the derivability problem has linear time complexity.
We prove the same complexity bound for ${\bf P}[\bot_w]$ in
Section \ref{DECISIONALGORITHM}.

Finally we consider a system ${\bf P}[\vp{}{}]$ which is the
modal-free fragment of Basic Propositional Primal Infon Logic {\bf
PPIL} from \cite{CotGur2012}. The primal disjunction $\vp{}{}$ in
${\bf P}[\vp{}{}]$ has usual introduction rules and no elimination
rules. We treat it as a group-key constructor and provide a linear
time reduction of ${\bf P}[\bot_w]$ to ${\bf P}[\vp{}{}]$. It thus
gives another proof of linear time complexity bound for ${\bf
P}[\bot_w]$.

\section{Semantics for $\{\top,\wedge,\imp{}{}\}$-fragment}\label{SEMANTICS}

Let $\Sigma$ be a
finite alphabet, say $\Sigma =\{0,1\}$. Let us fix a total pairing
function $\pi : \left ( \Sigma^*\right )^2\rightarrow \Sigma^*$
with projections $l,\,r : \Sigma^*\rightarrow \Sigma^*$, where
$\Sigma^*$ is the set of all binary strings,
\begin{equation}\label{PAIR}
l(\pi (x,y))=x,\quad r(\pi (x,y))=y,
\end{equation}
and two functions $\enc, \dec : \left ( \Sigma^*\right
)^2\rightarrow \Sigma^*$ such that $\enc$ is total and
\begin{equation}\label{ENC}
\dec (x, \enc(x,y))=y.
\end{equation}

String $\enc (x,y)$ will be treated as a ciphertext containing
string $y$ encrypted with key $x$. Function $\dec$ is the
decryption method that exploits the same key. In this text we do
not restrict ourselves to encryptions that  are strong in some
sense. For example, $\enc (x,y)$ may be the concatenation of
strings $x$ and $y$. Then $\dec$ on arguments $x,y$ simply removes
the prefix $x$ from $y$. The totality of functions $l$, $r$,
$\dec$ is not supposed, but the left-hand parts of (\ref{PAIR})
and (\ref{ENC}) must be defined for all $x,y\in \Sigma^*$.

We also fix some set $E\subset \Sigma^*$, $E\not =\emptyset$. It
will represent the information known by everyone, for example,
facts like $0<1$ and $2\cdot 2=4$. The structure ${\cal A}=\langle
\Sigma^*, \pi,l,r,\enc,dec,E\rangle$ will be referred as  {\em an
infon algebra}.\footnote{We use this term differently from
\cite{GurNee2009} where infon algebras are semi-lattices with
information order ``$x$ is at least as informative as $y$''. }

\begin{defin}\label{CLOSURECONDITIONS}\rm A set $M\subseteq \Sigma^*$
will be called {\em closed} if $E\subseteq M$ and $M$ satisfies
the following closure conditions:
\begin{enumerate}
\item $a,b\in M \Leftrightarrow \pi (a,b)\in M$,
\item $a, \enc
(a,b)\in M \Rightarrow b\in M$,
\item $a\in \Sigma^*,\,b\in M
\Rightarrow \enc(a,b)\in M$.
\end{enumerate}
\end{defin}

A closed set $M$ represents the information that is potentially
available to an agent in a local state, i.e. between two
consecutive communication steps of a protocol. The information is
represented by texts. $M$ contains all public and some private
texts. The agent can combine several texts in a single multi-part
document using $\pi$ function as well as to extract its parts by
means of $l$ and $r$. She has access to the encryption tool
$\enc$, so she can convert a plaintext into a ciphertext. The
backward conversion (by $\dec$) is also available provided she has
the encryption key.

Note that in the closure condition 3 we do not require that $a\in
M$. The agent will never need to decrypt the ciphertext
$\enc(a,b)$ encrypted by herself because she already has the
plaintext $b$. The key $a$ can be generated by some trusted third
party and sent to those who really need it. This is the case when
the encryption is used to provide secure communications between
agents when only the connections to the third party are secure
(and the authentication is reliable). On the other hand, some
protocols may require the agent to distribute keys by herself.
Then she can use a key that is known to her or get it from the
third party. In the latter case $a$ will be available in her new
local state that will be updated by the communication with the
third party.

\medskip
The natural deduction calculus for primal infon logic {\bf P} is
considered in \cite{BekGur2011}. The corresponding {\em
derivability relation} $\Gamma\vdash\varphi$ is defined by the
following rules:
$$
\ruleone{}{\vdash\top}\qquad
\ruleone{}{\varphi\vdash\varphi}\qquad
\ruleone{\Gamma\vdash\varphi}{\Gamma,\Delta\vdash\varphi\using{({\sl
Weakening})}}\qquad \ruletwo{ \Gamma\vdash\varphi_1 }{
\Gamma,\varphi_1\vdash\varphi_2 }{
\Gamma\vdash\varphi_2\using{({\sl Cut})} }
$$
$$
\ruletwo{\Gamma\vdash\varphi_1} {\Gamma\vdash\varphi_2}
{\Gamma\vdash\varphi_1\wedge \varphi_2\using{(\wedge I)}}\qquad
\ruleone{\Gamma\vdash\varphi_1\wedge \varphi_2}
{\Gamma\vdash\varphi_i\using{(\wedge E_i)\quad (i=1,2)}}
$$
$$
\ruleone{\Gamma\vdash \varphi_2}{\Gamma\vdash
\imp{\varphi_1}{\varphi_2} \using (\imp{}{}\!I)}
\qquad\ruletwo{\Gamma\vdash \varphi_1}{\Gamma\vdash
\imp{\varphi_1}{\varphi_2}}{\Gamma\vdash \varphi_2\using
(\imp{}{}\!E)} \;.
$$
Here $\varphi,\varphi_1,\varphi_2$ are infons, i.e. the
expressions constructed from the set $At$ of atomic infons  by the
 grammar
$$
\varphi ::= \top \mid At \mid (\varphi\wedge\varphi)\mid
(\imp{\varphi}{\varphi}),
$$
and $\Gamma$, $\Delta$ are sets of infons.

As usual, {\em a derivation of $\varphi$ from a set of assumptions
$\Gamma$} is a sequence of infons $\varphi_1,\ldots, \varphi_n$
where $\varphi_n=\varphi$ and each $\varphi_k$ is either a member
of $\Gamma\cup\{\top\}$ or is obtained from some members of
$\{\varphi_j\mid j<k\}$ by one of the rules
%$$
%\ruleone{\varphi\in \Gamma\cup\{\top\}}{\varphi}\qquad
%$$
$$
\ruletwo{\varphi_1} {\varphi_2} {\varphi_1\wedge
\varphi_2}\qquad \ruleone{\varphi_1\wedge \varphi_2}
{\varphi_i} \qquad \ruleone{ \varphi_2}{
\imp{\varphi_1}{\varphi_2} } \qquad\ruletwo{ \varphi_1}{
\imp{\varphi_1}{\varphi_2}}{ \varphi_2} \;.
$$
It is easy to see that $\Gamma\vdash\varphi$ iff there exists a
derivation of $\varphi$ from $\Gamma$. So rules like $({\sl
Weakening})$ or $({\sl Cut})$ from the definition of derivability
relation are never used in a derivation itself.

\begin{defin}\label{MODEL}\rm
{\em An interpretation}  (of the infon language) is a pair
$I=\langle{\cal A}, v \rangle$ where ${\cal A}=\langle \Sigma^*,
\pi,l,r,\enc,dec,E\rangle$ is an infon algebra and $v\!:At\cup \{
\top\}\rightarrow \Sigma^*$ is a total evaluation that assigns
binary strings  to atomic infons  and to constant $\top$,
$v(\top)\in E$. We assume that $v$ is extended as follows:
$$
v(\varphi_1\wedge \varphi_2)=\pi(v(\varphi_1), v(\varphi_2)),\quad
v(\imp{\varphi_1}{\varphi_2})=\enc(v(\varphi_1), v(\varphi_2)),
$$
$$
v(\Gamma)=\{v(\varphi)\mid \varphi \in\Gamma\}.
$$
{\em A model} is a pair $\langle I, M \rangle$ where $I$ is an
interpretation and $M\subseteq\Sigma^*$ is a closed set.

\end{defin}

In the paper \cite{BekGur2011} it is established that {\bf P} is
sound and complete with respect to quasi-boolean semantics. {\em A
quasi-boolean model} is a validity relation $\models$ that enjoys
the following properties:
\begin{itemize}
\item $\models\top$,
\item $\models \varphi_1\wedge \varphi_2
\;\Leftrightarrow\; \models \varphi_1 \mbox{ and }  \models
\varphi_2$,
\item $\models \varphi_2 \; \Rightarrow \; \models
\imp{\varphi_1}{\varphi_2}$,
\item $\models
\imp{\varphi_1}{\varphi_2}  \; \Rightarrow \; \not\models
\varphi_1 \mbox{ or }  \models \varphi_2$.
\end{itemize}
An infon $\varphi$ is derivable in the infon logic {\bf P}  from
the context $\Gamma$ iff $\models\Gamma$ implies $\models\varphi$
for all quasi-boolean models $\models$.

It can be seen that the definition of a quasi-boolean model is
essentially a special case of Definition \ref{MODEL}. Indeed,
suppose that atomic infons are words in the unary alphabet
$\{\mid\}$. Then all infons turn out to be words in some finite
alphabet $\Sigma_0$. Consider a translation $\ulcorner \cdot
\urcorner:\Sigma_0^*\rightarrow \{0,1\}^*$ that maps all elements
of $\Sigma_0$ into distinct binary strings of the same length,
$\ulcorner \Lambda \urcorner=\Lambda$ for the empty word $\Lambda$
and $ \ulcorner a_1\ldots a_n\urcorner = \ulcorner
\!\!a_1\!\!\urcorner\,\ldots \ulcorner\!\! a_n\!\!\urcorner $ \
for $a_1\ldots,a_n\in\Sigma_0$.

The corresponding infon algebra ${\cal A}$ and the evaluation $v$ can be defined
as follows: $v(a)=\ulcorner a \urcorner $ for $a\in At\cup \{\top\}$,
\begin{equation}\label{INEKTIVNO}
\pi(x,y)=\ulcorner ( \urcorner\, x \,\ulcorner\! \wedge
\!\urcorner \,y\, \ulcorner ) \urcorner\,,\quad \enc(x,y) =
\ulcorner ( \urcorner\, x \,\ulcorner \!\imp{}{}\! \urcorner \,y\,
\ulcorner ) \urcorner\, , \quad E=\{\ulcorner\top\urcorner\}.
\end{equation}
Projections and the decryption function can be found from
(\ref{PAIR}) and  (\ref{ENC}). Note that for this interpretation
the equality $v(\varphi)=\ulcorner
\varphi \urcorner $ holds for every infon $\varphi$.

Consider a quasi-boolean model $\models$. Let $M$ be the closure
of the set $M_0=\{\ulcorner \varphi \urcorner\mid \models \varphi
\}$, i.e. the least closed extension of $M_0$.

\begin{lem}\label{MODELLEMMA}
$\models \varphi$ \ iff \ $v(\varphi)\in M$.
\end{lem}

\proof It is sufficient to prove that the set $M\setminus M_0$
does not contain words of the form $v(\varphi)$. Any element $b\in
M\setminus M_0$ can be obtained from some elements of $M_0$ by a
finite sequence of steps 1,2,3 \ that correspond to closure
conditions:
\begin{enumerate}
\item $ x,y \mapsto \ulcorner ( \urcorner\, x \,\ulcorner \!\wedge
\! \urcorner \,y\, \ulcorner ) \urcorner ;\qquad \ulcorner (
\urcorner\, x \,\ulcorner \!\wedge
\! \urcorner \,y\, \ulcorner )
\urcorner \mapsto x $;\qquad $\ulcorner ( \urcorner\, x
\,\ulcorner \!\wedge
\! \urcorner \,y\, \ulcorner ) \urcorner \mapsto
y$; \item $ x,\;\ulcorner ( \urcorner\, x \,\ulcorner \!\imp{}{}\!
\urcorner \,y\, \ulcorner ) \urcorner \mapsto y; $ \item $ y
\mapsto \ulcorner ( \urcorner\, x \,\ulcorner \!\imp{}{}\!
\urcorner \,y\, \ulcorner ) \urcorner $.
\end{enumerate}

The history of this process is a derivation of $b$ from $M_0$ with
1,2,3 \ treated as inference rules. Let $b=v(\varphi)$ and
$b_1,\ldots,b_n=b$ be the derivation. Consider the (partial)
top-down syntactic analysis of strings $b_1,\ldots,b_n$ using
patterns
$$\ulcorner (
\urcorner\, \cdot \,\ulcorner \wedge
 \urcorner \,\cdot\ulcorner )
\urcorner \!,\!\qquad \ulcorner ( \urcorner\, \cdot \,\ulcorner
\!\imp{}{}\! \urcorner \,\cdot\, \ulcorner ) \urcorner,\qquad
\ulcorner \mid\mid\ldots \mid \urcorner.
$$
We replace all substrings that remain unparsed by $v(a)$ where
$a=||\ldots |$ is some fresh atomic infon. The resulting sequence
$c_1,\ldots,c_n$ is also a derivation of $b$ from $M_0$ because
any string of the from $v(\psi)$ has no unparsed substrings. All
its members have the form $c_i=v(\varphi_i)$ for some infons
$\varphi_i$. Moreover, $\varphi_1,\ldots,\varphi_n$ is a
derivation of $\varphi=\varphi_n$ in {\bf P} from the set of
hypotheses $\Gamma=\{\varphi_j\mid c_j\in M_0\}$. But $\models
\Gamma$ and {\bf P} is sound with respect to quasi-boolean models,
so $\models\varphi$ and $b=v(\varphi)\in M_0$. Contradiction. \fin

\begin{theo}\label{SOUNDANDCOMPLETE}
$\Gamma\vdash \varphi$ in {\bf P} iff \ $v(\varphi)\in M$ for
every model $\langle I, M \rangle$ with $v(\Gamma)\subseteq M$.
\end{theo}
\proof The theorem states that the infon logic {\bf P} is sound
and complete with respect to the class of models introduced by
Definition \ref{MODEL}. The soundness can be proven by
straightforward induction on the derivation of $\varphi$ from
$\Gamma$. The completeness follows from Lemma \ref{MODELLEMMA} and
the completeness result for quasi-boolean models (see
\cite{BekGur2011}).
 \fin

\bigskip A set  $\{v(\psi)\mid \psi\in T\}\subseteq \Sigma^*$ will
be called {\em deductively closed} if $T\vdash \psi$ implies $\psi
\in T$ for all infons $\psi$, i.e. $T$ is deductively closed in
{\bf P}. In the proof of Lemma \ref{MODELLEMMA} we actually
establish that the particular interpretation $\langle{\cal
A},v\rangle$ is {\em conservative} in the following sense: the
closure $M$ of any deductively closed set $M_0\subseteq\Sigma^*$
 does not contain ``new'' strings of the form
$v(\psi)\not \in M_0$. It is also {\em injective}:
$v(\varphi_1)=v(\varphi_2)$ implies $\varphi_1=\varphi_2$. An
interpretation that enjoys these two properties will be called
{\em plain}.

\begin{lem}
There exists a plain interpretation.
\end{lem}

The completeness part of Theorem \ref{SOUNDANDCOMPLETE} can be
strengthened.

\begin{theo}\label{STRONGCOMPLETENESS}
Let the interpretation $I=\langle{\cal A},v\rangle$ be plain. For
any context $\Gamma$ there exists a model $\langle I,M\rangle$
with $v(\Gamma)\subseteq M$ such that $\Gamma\not\vdash \varphi$
implies $v(\varphi)\not\in M$ for all infons $\varphi$.
\end{theo}

\proof Let $M$ be the closure of the set $M_0=\{v(\psi) \mid
\Gamma\vdash\psi\}$. Then $v(\Gamma)\subseteq M$. The set $M_0$ is
deductively closed, so $M\setminus M_0$ does not contain strings
of the form $v(\psi)$. Suppose $\Gamma\not\vdash \varphi$. Then
$v(\varphi)\not\in M_0$ because the interpretation is injective.
Thus $v(\varphi)\not\in M$. \fin

\section{Constant $\bot$ and backdoors} \label{CONSTANTBOT}
\subsection*{$\bot$ as superuser permissions}
Infon logic
${\bf P}[\bot]$ is the extension of ${\bf P}$ by additional
constant $\bot$ that satisfies the elimination rule
$$
\ruleone{\Gamma\vdash\bot}{\Gamma\vdash\varphi\using{(\bot E)}}\;.
$$

The corresponding changes in Definition \ref{MODEL} are as
follows. We add to the alphabet a new letter $\ff\not\in\Sigma$
and set $\Sigma_\bot = \Sigma\cup\{\ff\}$, \ $v(\bot)=\ff$.
Functions $\pi$, $l$, $r$, $\enc$, $\dec$  act on words from
$\Sigma_\bot^*$ but still satisfy the conditions (\ref{PAIR}),
(\ref{ENC}). We suppose them to preserve $\Sigma^*$: the value
should be a binary string provided all arguments are. We also
suppose that $v(\top)\in E\subseteq \Sigma^*$ and $v(\varphi)\in
\Sigma^*$ for $\varphi\in At$ and add new closure condition to
Definition \ref{CLOSURECONDITIONS}:

\medskip
4. $\ff\in M\,,a\in\Sigma_\bot^*\Rightarrow a\in M$.
\medskip

\noindent Models for ${\bf P}[\bot]$ are all pairs $\langle I,M
\rangle$ where $I$ is an interpretation and $M$ is a closed set,
both in the updated sense. The definition  of plain interpretation
is just the same.

Constant $\bot$ is some kind of root password that grants the superuser
permissions to its owner. The owner has the direct access to all
the information available in the system without any communication
or decryption. At the same time  $\bot$ can be incorporated into
some messages that will be used in communication.

%\par\medskip\noindent
%{\bf Example.} Assume that the general policy of an ordinary
%principal (not an administrator) is never to query the information
%that can be deduced by herself and to trust the administrator A on
%everything. Then A can switch any ordinary principal B off (make
%her keep silence) by sending her $\bot$. Indeed, after receiving
%$\bot$ principal B will know that \ $A \;said \;\bot$. She also
%knows that \ $A\; IsTrustedOn \;\bot$, \  so she derives
%$\bot$.\footnote{ Remember that $X\; IsTrustedOn \;y = \imp{(X
%\;said \;y)}{y}$. }  Now B can derive anything she wants by $(\bot
%E)$ rule and needs no communications at all.
%
%The administrator can also grant the ability to switch principal B
%off to some other principal C by sending \ $\imp{a}{\bot}$ \ and \
%$C\; IsTrustedOn \;a$ \ to B and $a$ to C. After receiving
%messages from A principals  B and C will derive the contents of
%the messages because they trust A. Any time later C may sent $a$
%to B and B will trust C on $a$. When it happens, B will derive $a$
%using \ $C\; IsTrustedOn \;a$ \ and finally derive $\bot$ using
%$\imp{a}{\bot}$. After it B will keep silence. \fin

\subsection*{$\bot$ as universal key}

The root password provides the direct access to all the
information in the system including private information of any
agent that was never sent to anybody else. It is also natural to
consider a restricted form of superuser permissions that protect
the privacy of agents but provide the ability to decrypt any
available ciphertext. It can be simulated by infon logic ${\bf
P}[\bot_w]$ with constant $\bot$ treated as {\em a universal key}.
The corresponding inference rule is a weak form of $(\bot E)$
rule,
$$
\ruletwo{ \Gamma\vdash \bot } { \Gamma\vdash\imp{\varphi}{\psi} }
{ \Gamma\vdash\psi \using{(\bot E_w)}, }
$$
that has an additional premise $\Gamma\vdash\imp{\varphi}{\psi}$.
So the owner of $\bot$ can get an infon only if she already has
the same information as a ciphertext. The rule $(\bot E_w)$ is
really weaker than $(\bot E)$ because $\imp{\psi}{\psi}$ is not
derivable in {\bf P}.

All definitions concerning models for ${\bf P}[\bot_w]$ are
similar to the case of ${\bf P}[\bot]$ with closure condition~4
replaced by

\medskip
$4'$. $\ff, \enc(a,b)\in M\Rightarrow b\in M$.
\medskip

\noindent Essentially we extend the signature of infon algebras by
additional (partial) operation $\crack(x,y)$ that satisfies the
equality
\begin{equation}\label{CRACK}
\crack(\ff,\enc(a,b))=b
\end{equation}
and allow any agent to use it, so her local state satisfies the
closure condition~$4'$.

\begin{lem}\label{PLAININT}
There exist plain interpretations for ${\bf P}[\bot]$ and for
${\bf P}[\bot_w]$.
\end{lem}
\proof We extend the example of plain interpretation for
$\{\top,\wedge,\imp{}{}\}$-fragment from Section \ref{SEMANTICS}
(see (\ref{INEKTIVNO})). Set $\ulcorner\bot\urcorner=\ff$ and
extend the interpretation in accordance with (\ref{INEKTIVNO}).
The resulting interpretation is plain in the sense of ${\bf
P}[\bot]$. Indeed, it is injective because $\ff\not\in\Sigma$. It
is also conservative. In order to prove this we use the
construction from Lemma \ref{MODELLEMMA}.

Let the set $M_0=\{v(\psi) \mid \psi\in T\}\subseteq
(\Sigma\cup\{\ff\})^*$  be deductively closed and $M$ be its
closure. Suppose $v(\varphi)\in M\setminus M_0$ for some infon
$\varphi$. Then $b_n=v(\varphi)$ has a derivation $b_1,\ldots,b_n$
from $M_0$ in the calculus with closure conditions considered as
inference rules:
\begin{enumerate}
\item $ x,y \mapsto \ulcorner ( \urcorner\, x \,\ulcorner \!\wedge
\!
\urcorner \,y\, \ulcorner ) \urcorner ;\qquad \ulcorner (
\urcorner\, x \,\ulcorner \!\wedge
\! \urcorner \,y\, \ulcorner )
\urcorner \mapsto x $;\qquad $\ulcorner ( \urcorner\, x
\,\ulcorner \!\wedge
\! \urcorner \,y\, \ulcorner ) \urcorner \mapsto
y$;
\item $ x,\;\ulcorner ( \urcorner\, x \,\ulcorner \!\imp{}{}\!
\urcorner \,y\, \ulcorner ) \urcorner \mapsto y; $
\item $ y
\mapsto \ulcorner ( \urcorner\, x \,\ulcorner \!\imp{}{}\!
\urcorner \,y\, \ulcorner ) \urcorner $;
\item $\ff \mapsto x$.
\end{enumerate}
Consider the (partial)
top-down syntactic analysis of strings $b_1,\ldots,b_n$ using
patterns
$$\ulcorner (
\urcorner\, \cdot \,\ulcorner\! \wedge \!
 \urcorner \,\cdot\ulcorner )
\urcorner \!,\!\qquad \ulcorner ( \urcorner\, \cdot \,\ulcorner
\!\imp{}{}\! \urcorner \,\cdot\, \ulcorner ) \urcorner,\qquad
\ulcorner \mid\mid\ldots \mid \urcorner,\qquad \ff.
$$
Replace all substrings that remain unparsed by $v(a)$ where
$a=||\ldots |$ is some fresh atomic infon. The resulting sequence
$c_1,\ldots,c_n$ is also a derivation of $v(\varphi)$ from $M_0$
because any string of the from $v(\psi)$ has no unparsed
substrings. All its members have the form $c_i=v(\varphi_i)$ for
some infons $\varphi_i$ and $\varphi_1,\ldots,\varphi_n$ is a
derivation of $\varphi=\varphi_n$ in ${\bf P}[\bot]$ from the set
of hypotheses $T$. But $T$ is deductively closed, so $\varphi\in
M_0$. Contradiction.

Now set
$$
\crack(x,y):= \left \{
\begin{array}{ll}
b, & \mbox{if $x=\ff$ and }y=\ulcorner
(\urcorner \,a\, \ulcorner \!\imp{}{}\!\urcorner\,b\, \ulcorner
)\urcorner, \\
\mbox{undefined,} & \mbox{otherwise.}
\end{array}
\right .
$$
It satisfies the condition (\ref{CRACK}), so the interpretation
for ${\bf P}[\bot_w]$ is defined. One can prove in a similar way
that the interpretation is plain (w.r.t. ${\bf P}[\bot_w]$ ).
\fin

The completeness results from Section \ref{SEMANTICS} hold for
logics ${\bf P}[\bot]$ and ${\bf P}[\bot_w]$ too. The proofs are
essentially the same with one difference: the quasi-boolean
semantics from \cite{BekGur2011} does not cover the case of ${\bf
P}[\bot_w]$. Let {\bf L} be one of the logics ${\bf P}[\bot]$ or
${\bf P}[\bot_w]$.

\begin{theo}\label{COMPLETENESSBOTTOM}
$\Gamma\vdash \varphi$ in {\bf L} iff \ $v(\varphi)\in M$ for
every model $\langle I, M \rangle$ of {\bf L} with
$v(\Gamma)\subseteq M$.
\end{theo}
\proof The soundness part can be proven by straightforward
induction on the derivation of $\varphi$ from $\Gamma$. The
completeness follows from Lemma \ref{PLAININT} and Theorem
\ref{STRONGCOMPLETENESSBOTTOM}. \fin

\begin{theo}\label{STRONGCOMPLETENESSBOTTOM}
Let $I$ be a plain interpretation of {\bf L}. For any context
$\Gamma$ there exists a model $\langle I,M\rangle$ of {\bf L} with
$v(\Gamma)\subseteq M$ such that $\Gamma\not\vdash \varphi$
implies $v(\varphi)\not\in M$ for all infons $\varphi$.
\end{theo}
\proof Similar to Theorem \ref{STRONGCOMPLETENESS}. \fin

\section{Decision algorithm for ${\bf P}[\bot_w]$}\label{DECISIONALGORITHM}

The derivability problems for infon logics ${\bf P}$ and ${\bf
P}[\bot]$ are linear time decidable (\cite{GurNee2011},
\cite{BekGur2011}, \cite{CotGur2012}). We provide a decision
algorithm for ${\bf P}[\bot_w]$ with the same complexity bound.

\medskip
\begin{defin}\label{POSITIVEAT}\rm (Positive atoms.)
In what follows we assume that the language of ${\bf P}$ also
contains $\bot$, but it is an ordinary member of $At$ without any
specific inference rule for it. Let
$$
\begin{array}{l}
At^+(\varphi)= \{\varphi \}  \mbox{ for } \varphi\in
At\cup\{\top,\bot\},\\[3pt]
At^+(\varphi\wedge\psi)=At^+(\varphi)\cup At^+(\psi),  \\[3pt]
At^+(\imp{\varphi}{\psi})=At^+(\psi).
\end{array}
$$
For a context $\Gamma$ set
$At^+(\Gamma)=\bigcup_{\varphi\in\Gamma}At^+(\varphi)$.
\end{defin}

\begin{lem}\label{LEMMACORTWO}
Let $\Gamma\vdash\bot$ in ${\bf P}[\bot_w]$. Then
$\Gamma\vdash\varphi$ in ${\bf P}[\bot_w]$ iff
$At^+(\varphi)\subseteq At^+(\Gamma)$.
\end{lem}
\proof Suppose $\Gamma\vdash\varphi$. The inclusion
$At^+(\varphi)\subseteq At^+(\Gamma)$ can be proved by
straightforward induction on the derivation of $\varphi$ from
$\Gamma$.

Now suppose that $\Gamma\vdash\bot$ and $At^+(\varphi)\subseteq
At^+(\Gamma)$. By rules $(\wedge\,E_i)$ and $(\bot E_w)$ we prove
that $\Gamma\vdash\psi$ for every infon $\psi\in At^+(\Gamma)$.
Then we derive $\Gamma\vdash\varphi$ by rules $(\wedge\,I)$,
$(\imp{}{}I)$. \fin

\begin{lem}\label{LEMMACORONE}
If  $\Gamma\not\vdash\bot$ in ${\bf P}$ and $\Gamma\vdash\varphi$
in ${\bf P}[\bot_w]$ then $\Gamma\vdash\varphi$ in ${\bf P}$.
\end{lem}
\proof $\Gamma\not\vdash\bot$ in ${\bf P}$ implies that
$\Gamma\not\vdash\bot$ in ${\bf P}[\bot_w]$ because the shortest
derivation of $\bot$ from $\Gamma$ cannot use the $(\bot E_w)$
rule. So any derivation in ${\bf P}[\bot_w]$ from $\Gamma$ cannot
use this rule. \fin

The decision algorithm for ${\bf P}[\bot_w]$ consists of the
following three steps:

\begin{enumerate}
\item Test whether $\Gamma\vdash\varphi$ in ${\bf P}$. If yes,
then $\Gamma\vdash\varphi$ in ${\bf P}[\bot_w]$ too. Else go to
step~2.

\item Test whether $\Gamma\not\vdash\bot$ in ${\bf P}$. If yes,
then $\Gamma\not\vdash\varphi$ in ${\bf P}[\bot_w]$ by Lemma
\ref{LEMMACORONE}. Else go to step~3.

\item We have $\Gamma\vdash\bot$ in ${\bf P}$, so it is also true
in ${\bf P}[\bot_w]$. Test the condition $At^+(\varphi)\subseteq
At^+(\Gamma)$. If it is fulfilled then $\Gamma\vdash\varphi$ in
${\bf P}[\bot_w]$; otherwise $\Gamma\not\vdash\varphi$ in ${\bf
P}[\bot_w]$ (Lemma \ref{LEMMACORTWO}).

\end{enumerate}

Linear time complexity bounds for steps 1,2 follow from the linear
bound for {\bf P}. In order to prove the same bound for step 3 we
use the preprocessing stage of the linear time decision algorithm
from \cite{CotGur2012}. It deals with sequents
$\Gamma\vdash\varphi$ in a language that extends the language of
${\bf P}[\bot_w]$. The preprocessing stage is purely syntactic, so
it does not depend on the logic involved and can be used for ${\bf
P}[\bot_w]$ as well.

The algorithm constructs the parse tree for the sequent. Two nodes
are called homonyms if they represent two occurrences of the same
infon. For every homonymy class, the algorithm chooses a single
element of it, the homonymy leader, and labels all nodes with
pointers that provide a constant time access from a node to its
homonymy leader. All this can be done in linear time (see
\cite{CotGur2012}).

Now it takes a single walk through the parse tree to mark by a
special flag all homonymy leaders that correspond to infons
$\psi\in At^+(\Gamma)$. One more walk is required to test whether
all homonymy leaders that correspond to $\psi\in At^+(\varphi)$
already have this flag. Thus we have a linear time test for the
inclusion $At^+(\varphi)\subseteq At^+(\Gamma)$.

\begin{theo}\label{LINEARTIME}
The derivability problem for infon logic ${\bf P}[\bot_w]$ is
linear time decidable.
\end{theo}

\section{Primal disjunction and backdoor emulation}

Primal infon logic with disjunction  ${\bf P}[\vee]$ was studied
in \cite{BekGur2011}. It is defined by all rules of {\bf P} and
usual introduction and elimination rules for disjunction. ${\bf
P}[\vee]$ can emulate the classical propositional logic, so the
derivability problem for it is co-NP-complete.

Here we consider the logic ${\bf P}[\vp{}{}]$, an efficient
variant of ${\bf P}[\vee]$. It was mentioned in \cite{BekGur2011}
and later was incorporated into  Basic Propositional Primal Infon
Logic {\bf PPIL} \cite{CotGur2012} as its purely propositional
fragment without modalities. In ${\bf P}[\vp{}{}]$ the standard
disjunction is replaced by a ``primal'' disjunction $\vp{}{}$ with
introduction rules
$$
\ruleone{\Gamma\vdash \varphi_i}{\Gamma\vdash
\vp{\varphi_1}{\varphi_2}\using{(\vp{}{}I_i)}} \qquad (i=1,2)
$$
and without elimination rules. It results in a linear-time
complexity bound for ${\bf P}[\vp{}{}]$ (and for {\bf PPIL} too,
see \cite{BekGur2011},\cite{CotGur2012}).

When the primal implication is treated as encryption, the primal
disjunction can be used as a method to construct group keys. An
infon of the form
\begin{equation}\label{GROUPKEY}
\imp{(\vp{\varphi_1}{\varphi_2})\;}{\;\psi}
\end{equation}
represents a ciphertext that can be decrypted by anyone  who has
at least one of the keys $\varphi_1$ or $\varphi_2$. In {\bf P}
the same effect can be produced by the infon
\begin{equation}\label{GROUPKEYEMULATION}
(\imp{\varphi_1}{\psi})\wedge (\imp{\varphi_2}{\psi}),
\end{equation}
but it requires two copies of $\psi$ to be encrypted. Moreover, a
principal A who does not know both keys $\varphi_1$ and
$\varphi_2$ fails to distinguish between (\ref{GROUPKEYEMULATION})
and $ (\imp{\varphi_1}{\psi_1})\wedge (\imp{\varphi_2}{\psi_2})$.
If A receives (\ref{GROUPKEYEMULATION}) from some third party and
forwards it to some principals B and C, she will never be sure
that  B and C will get the same plaintext after decryption. Group
keys eliminate the length growth and ambiguity.

An infon algebra for ${\bf P}[\vp{}{}]$ has an additional total
operation $\gr \!: (\Sigma^*)^2\rightarrow \Sigma^*$ for
evaluation of primal disjunction: $ v(\vp{\varphi}{\psi}) =
\gr(v(\varphi),v(\psi))$. The corresponding closure condition in
Definition \ref{CLOSURECONDITIONS} will be

\medskip
5. If $a\in M,\,b\in\Sigma^*$ or $b\in M,\,a\in\Sigma^*$ then $\gr
(a,b)\in M$.

\medskip\noindent
All the results of Section \ref{CONSTANTBOT} (Lemma
\ref{PLAININT}, Theorems \ref{COMPLETENESSBOTTOM},
\ref{STRONGCOMPLETENESSBOTTOM}) hold for ${\bf P}[\vp{}{}]$ too.
The proofs are essentially the same.

\paragraph{
${\bf P}[\bot_w]$ is linear-time reducible to ${\bf P}[\vp{}{}]$,}
so ${\bf P}[\vp{}{}]$ and {\bf PPIL} can emulate the backdoor
based on a universal key. The reduction also gives another proof
for Theorem \ref{LINEARTIME}.

Remember that in the language of ${\bf P}[\vp{}{}]$ symbol $\bot$
denotes some regular atomic infon. Consider the following
translation:
$$
\begin{array}{l}
q^*=q \mbox{ \ for }q\in At\cup\{\top,\bot\}, \\[3pt]
(\varphi\wedge\psi)^*= \varphi^*\wedge\psi^*,\\[3pt]
(\imp{\varphi}{\psi})^*=\imp{(\vp{\bot}{\varphi^*})\;}{\;\psi^*},\\[3pt]
\Gamma^*=\{\varphi^*\mid \varphi\in\Gamma\}.
\end{array}
$$
The transformation of $\Gamma,\varphi$ into $\Gamma^*,\varphi^*$
can be implemented in linear time.

\begin{theo}
$\Gamma\vdash\varphi$ in ${\bf P}[\bot_w]$ iff \
$\Gamma^*\vdash\varphi^*$ in ${\bf P}[\vp{}{}]$.
\end{theo}
\proof Part ``only if'' can be proved by straightforward induction on
the derivation of $\varphi$ from assumptions $\Gamma$ in ${\bf
P}[\bot_w]$. For any inference rule of ${\bf P}[\bot_w]$, its
translation is derivable in ${\bf P}[\vp{}{}]$. For example,
consider the elimination rules for $\imp{}{}$ and $\bot$:
$$
\ruletwo{
\ruleone{\varphi^*}
{\vp{\bot}{\varphi^*}}
}
{
\imp{\vp{\bot}{\varphi^*}\;}{\;\psi^*}
}
{
\psi^*\using{\;,}
}
\qquad\qquad
\ruletwo{
\ruleone{\bot}
{\vp{\bot}{\varphi^*}}
}
{
\imp{\vp{\bot}{\varphi^*}\;}{\;\psi^*}
}
{
\psi^* \using{\;.}
}
$$

Part ``if''. Let $\Gamma^*\vdash\varphi^*$ in ${\bf P}[\vp{}{}]$.
Note that ${\bf P}[\vp{}{}]$ is the modal-free fragment of {\bf
PPIL} and the shortest derivation of $\varphi^*$ from assumptions
$\Gamma^*$ in {\bf PPIL} is also a derivation in ${\bf
P}[\vp{}{}]$. Let $D$ be this derivation.

It is proved in \cite{CotGur2012} that any shortest derivation is
local. For the case of ${\bf P}[\vp{}{}]$ it means that all
formulas from $D$ are subformulas of $\Gamma^*, \varphi^*$. In
particular, $\vp{}{}$ occurs in $D$ only in subformulas of the
form $\vp{\bot}{\theta^*}$.

Case 1. Suppose that the $(\vp{}{}I_1)$ rule is never used in $D$.
Remove part ``$\vp{\bot}{}$'' from every subformula of the form
$\vp{\bot}{\psi}$ that occurs in $D$. The result will be a
derivation of $\varphi$ from assumptions $\Gamma$ in {\bf P}. So
$\Gamma\vdash\varphi$ in ${\bf P}[\bot_w]$ too.

Case 2. Suppose that the $(\vp{}{}I_1)$ rule is used in $D$. It has
the form
\begin{equation}\label{BOTORRULE}
\ruleone{\bot}{\vp{\bot}{\theta^*}}\;,
\end{equation}
so $D$ also contains a derivation of $\bot$. The corresponding
subderivation is the shortest one and does not use the $(\vp{}{}I_1)$
rule. By applying the transformation from Case 1 we prove that
$\Gamma\vdash\bot$ in {\bf P} and $\bot\in At^+(\Gamma)$.

We extend Definition \ref{POSITIVEAT} with new item
$$
At^+(\vp{\psi_1}{\psi_2})=At^+(\psi_1)\cup At^+(\psi_2),
$$
so $At^+(\psi)$ is defined for every $\psi$ in the language of
${\bf P}[\vp{}{}]$. Moreover, $At^+(\varphi^*)=At^+(\varphi)$ and
$At^+(\Gamma^*)=At^+(\Gamma)$. We claim that
$At^+(\varphi^*)\subseteq At^+(\Gamma^*)$.

Indeed, consider $D$ as a proof tree and its node  $\psi$ with
$At^+(\psi)\not\subseteq At^+(\Gamma^*)$ whereas
$At^+(\psi')\subseteq At^+(\Gamma^*)$ holds for all predecessors
$\psi'$. The only rule that can produce this effect is
(\ref{BOTORRULE}), so $\psi=\vp{\bot}{\theta^*}$ for some $\theta$
where all occurrences of ``new'' atoms $q\in At^+(\psi)\setminus
At^+(\Gamma^*)$ are inside $\theta^*$.

Consider the path from the node $\psi$ to the root node
$\varphi^*$ and the trace of $\psi$ along it. There is no
elimination rule for $\vp{}{}$, so $\psi$ cannot be broken into
pieces. All occurrences of positive atoms in $\theta^*$ will be
positive in all formulas along the trace. But $\vp{}{}$ occurs in
$\varphi^*$ only in the premise of primal implication, so the
trace does not reach the root node. Thus, at some step the formula
containing $\psi$ will be eliminated and ``new'' atoms from
$\theta^*$ will never appear in $At^+(\varphi^*)$:
$$
\ruletwo{
\ruleone{
\ruleone{\bot}{\vp{\bot}{\theta^*}}
}
{\ruledot{\eta_1[\vp{\bot}{\theta^*}]}}
}
{\ruledot{\imp{\eta_1[\vp{\bot}{\theta^*}]}{\eta_2}}}
{\eta_2}
$$

We have established that $At^+(\varphi)\subseteq At^+(\Gamma)$.
But $\Gamma\vdash\bot$ in {\bf P} and in ${\bf P}[\bot_w]$, so
$\Gamma\vdash\varphi$ in ${\bf P}[\bot_w]$ by Lemma
\ref{LEMMACORTWO}. \fin

\com It is also possible to reduce ${\bf P}[\bot_w]$ to {\bf P}.
The corresponding reduction is two-step translation. One should
convert $\varphi$ into $\varphi^*$ and then replace all
subformulas of the form (\ref{GROUPKEY}) in it with
(\ref{GROUPKEYEMULATION}). Unfortunately, the second step results
in the exponential growth of the length of a formula.

\section*{Acknowledgements}

I would like to thank Yuri Gurevich, Andreas Blass and Lev Beklemishev for
valuable discussion, comments and suggestions.

The research described in this paper was partially supported by
Microsoft project DKAL and Russian Foundation for Basic Research
(grant 11-01-00281).

\end{document}